\documentclass[a4paper,
               keeplastbox,   
               ]{jacow}
\usepackage{pdfpages,multirow,ragged2e} %
\makeatletter%
	\ifboolexpr{bool{xetex}}
	 {\renewcommand{\Gin@extensions}{.pdf,%
	                    .png,.jpg,.bmp,.pict,.tif,.psd,.mac,.sga,.tga,.gif,%
	                    .eps,.ps,%
	                    }}{}
\makeatother

\ifboolexpr{bool{xetex} or bool{luatex}} 
 {}                                      
 {\usepackage[utf8]{inputenc}}           

\usepackage[USenglish]{babel}

\ifboolexpr{bool{jacowbiblatex}}%
 {%
  \addbibresource{jacow-test.bib}
  \addbibresource{biblatex-examples.bib}
 }{}
\begin{document}

\setlength{\abovedisplayskip}{3pt}
\setlength{\belowdisplayskip}{3pt}

\title{Narrow bandwidth Active Noise Control for microphonics rejection in superconducting cavities at \NoCaseChange{LCLS-II} \thanks{Work supported by the DOE and Helmholtz Institute}}

\author{A. Bellandi \thanks{andrea.bellandi@desy.de}, J. Branlard, \\ Deutsches Elektronen-Synchrotron DESY, Notkestr. 85, 22607 Hamburg, Germany \\
		J. Diaz Cruz\textsuperscript{1}, S. Aderhold, A. Benwell, A. Brachmann, S. Hoobler, A. Ratti,  \\
		D. Gonnella, J. Nelson, R. D. Porter, L. Zacarias, \\ 
		SLAC, Menlo Park, CA 94025, USA\\
		\textsuperscript{1}also at University of New Mexico, Albuquerque, NM 87131, USA}
	
\maketitle

\begin{abstract}
   LCLS-II is an X-Ray Free Electron Laser (XFEL) commissioned in 2022, being the first Continuous Wave (CW) hard XFEL in the world to come into operation. To accelerate the electron beam to an energy of $\SI{4}{\giga \eV}$, 280 TESLA type superconducting RF (SRF) cavities are used. A loaded quality factor ($Q_L$) of $4 \times 10^7$ is used to drive the cavities at a power level of a few kilowatts. For this $Q_L$, the RF cavity bandwidth is 32 Hz. Therefore, keeping the cavity resonance frequency within such bandwidth is imperative to avoid a significant increase in the required drive power. In superconducting accelerators, resonance frequency variations are produced by mechanical microphonic vibrations of the cavities. One source of microphonic noise is rotary machinery such as vacuum pumps or HVAC equipment. A possible method to reject these disturbances is to use Narrowband Active Noise Control (NANC) techniques. These techniques were already tested at DESY/CMTB \cite{DESY} and Cornell/CBETA \cite{CORNELL}. This proceeding presents the implementation of a NANC controller adapted to the LCLS-II Low Level RF (LLRF) control system. Tests showing the rejection of LCLS-II microphonic disturbances are also presented.
\end{abstract}

\section{INTRODUCTION}
   The SRF cavity is the device responsible for storing energy in the form of electromagnetic field in a particle accelerator. When a particle beam passes through a cavity, it interacts with the field by exchanging energy with it. An SRF cavity can be modeled as a narrow band filter centered around its resonance frequency $f_0$. Even though the nominal resonance frequency is specified at the cavity design stage, it can change due to unwanted mechanical deformations during operation. The difference between the nominal resonance frequency and the actual one is referred as \textit{detuning}, $\Delta f$.
   
   The relation between power consumption and detuning of a steady state cavity when driving it at its design resonance frequency is
   
   \begin{align}
       P_f &= \frac{V_c^2}{4\frac{R}{Q}Q_L}[1 + \left( \frac{\Delta f}{f_{1/2}} \right)^2], \\
       Q_L &= \frac{f_0}{2 f_{1/2}},
   \end{align}
   
   where $P_f$ is the required power to drive a cavity at the accelerating voltage $V_c$, $f_{1/2}$ is the cavity half bandwidth and $\frac{R}{Q}$ is the cavity shunt impedance. Therefore, even if decreasing $f_{1/2}$ can reduce the power consumption needed to reach a certain accelerating gradient, detuning signals that are comparable or higher than $f_{1/2}$ can negate such an advantage. 
   
   Resonance control systems then play a crucial role in modern SRF CW accelerators for FELs, by keeping the cavities close to the nominal resonance frequency. For LCLS-II the maximum peak detuning that can be withstand without lowering the gradient due to power constraints is equal to $\pm\SI{10}{\hertz}$. A method to reduce the cavity detuning is to provide active compensation using the piezoelectric tuner actuators attached to the cavity. In LCLS-II, these tuners can steer the cavity frequency by up to $\SI{2}{\kilo \hertz}$.
   
\section{TYPES OF MICROPHONICS DISTURBANCES}
   The cavity mechanical disturbances have different spectral and time characteristics. However, two kind of external mechanical perturbations are commonly found in every CW SRF system and are accountable for the major part of the cavity detuning.
   \begin{itemize}
       \item Drifts produced by the cooling system. Since SRF cavities are located in tanks filled with super-fluid cryogenic helium, every drift of the pressure results in a detuning variation. The typical timescale of this effect is usually higher than seconds.  
       \item Vibrations produced by machinery located in the proximity of the accelerator. Due to the necessity of actively maintaining some physical characteristics in the accelerating system, like high vacuum and cryogenic temperatures, different kinds of pumps are required. This kind of noise has a narrow bandwidth and a sinusoidal-like shape.
   \end{itemize}
   
   


\section{DESCRIPTION AND IMPLEMENTATION OF THE NANC}
 For drifts generated by the cooling system, a controller with an integral policy is generally used, and it is not discussed in this paper. For microphonics vibrations generated by machinery, NANC\cite{kuo1999active} techniques were demonstrated in the last decade at CMTB \cite{DESY} and CBETA \cite{CORNELL}. The advantage of this kind of controller is the capability of rejecting sinuisoidal disturbances without exciting mechanical modes of the cavity. Additionally, NANC is an adaptive controller, meaning that it is capable of self-correct its internal parameters against variations of the noise properties. Therefore such a controller was chosen to test microphonics rejection on LCLS-II cavities.
 
 The principle of operation of the NANC is to generate a sinusoidal signal to drive the piezo actuator with the same frequency as the microphonic detuning source $\omega_m$, but with $180^{\circ}$ of phase offset; this way, the detuning source will be perfectly cancelled. In principle, one should have a sinusoidal component for every source of microphonics. As described in \cite{CORNELL}, the control signal $u_{pzt}$ that drives the piezoelectric tuners is the superposition of multiple In-phase and Quadrature ($I_m$ and $Q_m$) components at different microphonics frequencies $\omega_m$:

 \begin{equation}\label{eq:piez_voltage}
    u_{\rm{pzt}}(t)=\sum_{m}{I_m(t)cos(\omega_m(t)t)+Q_m(t)sin(\omega_m(t)t)}
 \end{equation}

 where the subindex $m$ represents each component of the microphonics detuning spectrum. The NANC controller is based on the gradient descent algorithm, and it updates the $I_m$ and $Q_m$ values at each iteration using the present values at $t_n$ and the adaptation rate $\mu_m$

 \begin{subequations}\label{eq:I_Q_update}
    \begin{align}
    I_m(t_{n+1})=I_m(t_n)-\mu_m\Delta f(t_n) \times \nonumber\\ cos(\omega_m(t)t_n-\phi_m(t_n)),
    \label{eq:I_update}
    \\
    Q_m(t_{n+1})=Q_m(t_n)-\mu_m\Delta f(t_n) \times \nonumber\\ sin(\omega_m(t)t_n-\phi_m(t_n)),
    \label{eq:Q_update}
    \end{align}
 \end{subequations}

 In addition, $\phi_m$ represents the lag of the tuner at a given frequency, and it is also updated at each iteration by the algorithm using Equation \ref{eq:phi_update}, where $\eta_m$ is the adaptation rate.

 \begin{equation}\label{eq:phi_update}
 \begin{split}
    \phi_m(t_{n+1}) = \; \phi_m(t_n) - &\eta_m\Delta f(t_n) \times \\ \{ &I_m(t_n)\sin(\omega_m(t)t_n-\phi_m(t_n)) -\\ 
    &Q_m(t_n)\cos(\omega_m(t)t_n-\phi_m(t_n))\}
 \end{split} 
 \end{equation}
 
 In addition to amplitude and phase delay adaptation, we have implemented a frequency follower feature that tracks the exact frequency of the microphonics source, making the controller robust to drifts on the microphonics frequency. 

 \begin{equation}\label{eq:freq_update}
 \begin{split}
    \omega_m(t_{n+1}) = \; &\omega_m(t_{n}) + \\
    &\gamma_m \frac{I_m(t_n)Q_m(t_{n+1}) - Q_m(t_n)I_m(t_{n+1})}{I_m(t_n)^2 + Q_m(t_n)^2}
 \end{split}
 \end{equation}
 
 where $\gamma_m$ is the frequency adaptation rate.
 
\begin{figure*}[h]
    \centering
    \includegraphics*[width=0.8\textwidth]{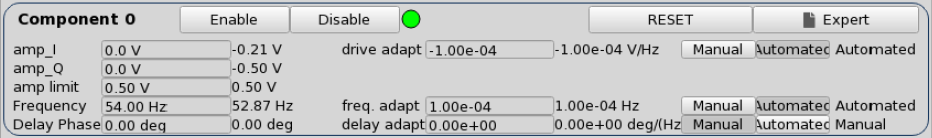}
    \caption{NANC Controller user interface for an individual component.}
    \label{fig:nanc_pane}
\end{figure*}

 To use the controller a PyDM\cite{PYDM} user interface (UI) was developed (Fig. \ref{fig:nanc_pane}). From the UI, the amplitude of $I_m$ and $Q_m$, the phase delay $\phi_m$, and the frequency $\omega_m$ can be modified manually or automatically, using the NANC algorithm. When the NANC algorithm is enabled, the user should set the values of the adaptation rates $\mu_m$, $\eta_m$ and $\gamma_m$. Up to four frequency components can be enabled to drive the piezoelectric actuator. 
 
 The NANC controller was implemented in the Resonance Control Chassis' (RCC) FPGA of the LCLS-II LLRF system \cite{LCLS-II LLRF}, using one soft CPU per cavity. The choice of using a soft CPU instead of using directly the programmable logic of the FPGA was done to speed-up the algorithm development. Such a choice is made possible due to the frequency of the microphonic disturbances, in the order of tens to hundreds of Hertz. Therefore a sample rate of $\SI{5548}{\hertz}$, which is calculated by decimating the computed detuning sample stream by 1024 using a Cascaded Comb-Integrator\cite{CIC} filter of the first order, is deemed to be enough to compensate for the microphonic disturbances. At the same time, such a value for the sample rate gives enough time to the NANC algorithm to complete one iteration before the arrival of the next detuning sample. The final occupation of the RCC's FPGA with four soft CPUs is listed in Table~\ref{tab:resources}. The occupation is compared to the original resonance controller project which includes a resonance controller written by the Fermi National Accelerator Laboratory (FNAL)\cite{einstein2017,holzbauer2018}.
 
 \setlength{\tabcolsep}{1pt} 
     \begin{table}[hbt!]
   \centering
   \caption{Resource occupation of the NANC and FNAL ANC resonance controller on Xilinx XC7K160T FPGA.}
   \begin{tabular}{lccccc}
       \toprule
        & \multicolumn{2}{c}{\textbf{NANC}} & \multicolumn{2}{c}{\textbf{FNAL ANC}} & \textbf{Available}\\
        & Used & Used [\%] & Used & Used [\%]\\
       \midrule
        LUT & 45202 & 44.58 & 34354 & 33.88 & 101400\\
        Flip Flop & 50300 & 24.8 & 48617 & 23.97 & 202800\\
        Block RAM & 171 & 52.62 & 96.5 & 29.69 & 325\\
        DSP slices & 64 & 10.67 & 148 & 24.67 & 600\\
       \bottomrule
   \end{tabular}
   \label{tab:resources}
\end{table}

\section{TEST OF THE CONTROLLER}
\begin{figure}[h]
\centering
   \includegraphics[width=0.45\textwidth]{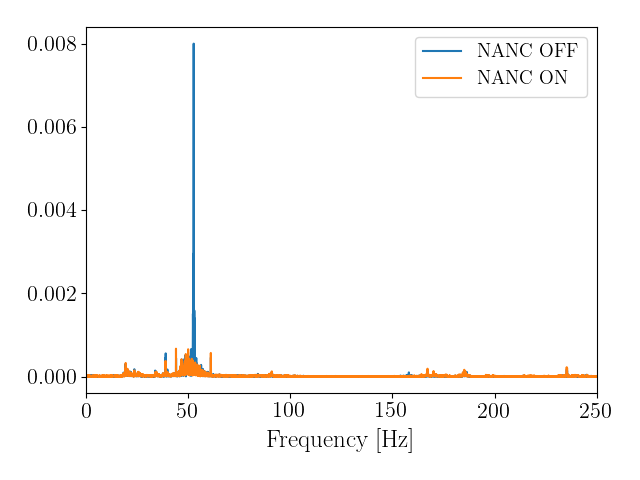}
   \caption{Detuning spectrum of CM16 cavity 1.}
   \label{fig:CM16_cav1_spectrum}
\end{figure}

The controller was tested on the SRF cavities of the LCLS-II LINAC at a temperature of 2 K and with a gradient of 5 MV. It was tested with the field control both in open and closed loop. Here we present results for just one of the SRF cavities, as these results are representative of the performance of the controller. For cavity 1 of cryomodule (CM) 16, initial microphonics measurements showed a detuning component at ~$\SI{54}{\hertz}$, as shown by the blue curve of Fig. \ref{fig:CM16_cav1_spectrum}; therefore, we enabled one component of the NANC controller, setting its frequency to $\SI{54}{\hertz}$.

\begin{figure}[h]
\centering
   \includegraphics[width=0.45\textwidth]{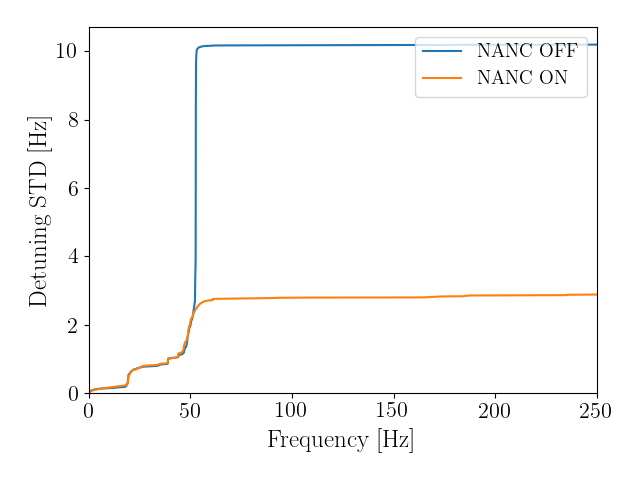}
   \caption{Cumulative detuning standard deviation on CM16 cavity 1.}
   \label{fig:CM16_cav1_detuning_std}
\end{figure}

\begin{figure}[h]
\centering
   \includegraphics[width=0.45\textwidth]{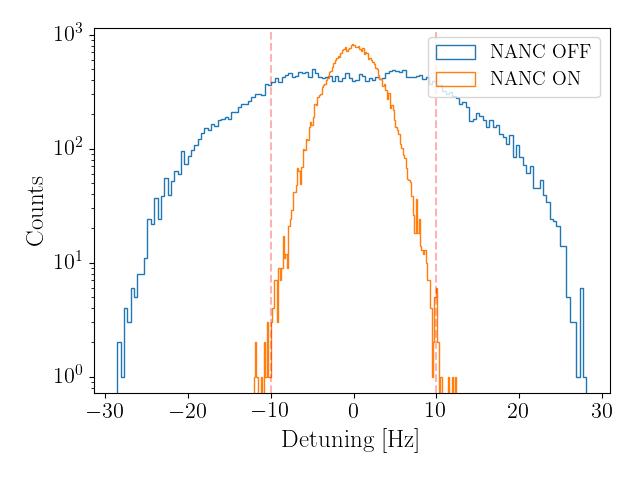}
   \caption{Detuning distribution of CM16 cavity 1. The red dashed lines indicates the LCLS-2 specification limits for detuning.}
   \label{fig:CM16_cav1_dist}
\end{figure}

\begin{figure}[h]
\centering
   \includegraphics[width=0.45\textwidth]{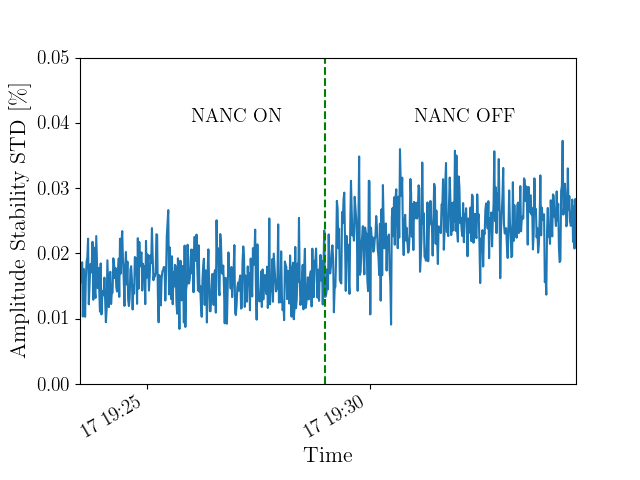}
   \caption{Amplitude stability of CM16 cavity 1. the vertical green dotted line indicates when the NANC tool was disabled.}
   \label{fig:arstd}
\end{figure}

The effect of the controller in the cavity detuning can be seen in Fig.~ \ref{fig:CM16_cav1_spectrum}, \ref{fig:CM16_cav1_detuning_std} and \ref{fig:CM16_cav1_dist}. The controller successfully compensates for narrowband microphonics, as seen in Fig \ref{fig:CM16_cav1_spectrum}, where the main microphonics component vanishes when NANC is enabled. From Fig.~\ref{fig:CM16_cav1_detuning_std}, the reduction in cavity detuning is $\sim9$ dB, and from Fig. \ref{fig:CM16_cav1_dist} the cavity detuning lays in between detuning requirements for LCLS-II when NANC is enabled. Besides confirming the reduction of the detuning noise, the impact on field stability was assessed. Fig. \ref{fig:arstd} shows the amplitude stability standard deviation with NANC enabled and disabled, and one can see an improvement of $\sim0.01\%$ when the tool is enabled. 

\section{CONCLUSIONS}
The NANC previously implemented at DESY and CORNELL has now been implemented at SLAC for the LCLS-II LINAC in CW operation, with the addition of the frequency follower feature. The controller is able to compensate for narrowband microphonics and reduces cavity detuning by a factor of 3. With this microphonics compensation, detuning of noisy cavities is reduced and reaches LCLS-II specifications. More cavities should be tested to have a better understanding of the performance of the controller.  

\section{ACKNOWLEDGEMENTS}
The authors would like to acknowledge the entire LCLS-II LLRF team at SLAC and partner labs and the SRF team at SLAC who kindly provided their support to test the controller. The authors also acknowledge European XFEL GmbH for funding the scientific collaboration between SLAC and DESY.
%
%
\ifboolexpr{bool{jacowbiblatex}}%
	{\printbibliography}%
	{%
	

} 
%
%


\end{document}